\documentclass[preprint,pre,eqsecnum,showpacs,amsmath,%
amssymb,nofootinbib,showkeys,byrevtex,aps]{revtex4}
\usepackage{graphicx}
\usepackage{amsmath}
\usepackage{bm}
\newcommand{\be}{\begin{equation}}
\newcommand{\ee}{\end{equation}}
\begin{document}
\title{Quasi-normal modes of a dielectric sphere \\ and some their implications}
\author{V.~V.~Nesterenko}
\affiliation{Bogoliubov Laboratory of Theoretical Physics, Joint
Institute for Nuclear Research, 141980 Dubna, Russia}
\email{nestr@theor.jinr.ru}
\author{A.~Feoli}
\affiliation{Dipartimento di Ingegneria, Universit\`{a} del
Sannio, Corso Garibaldi n.\ 107, Palazzo Bosco Lucarelli,  82100
Benevento,  Italy} \affiliation{INFN Sezione di Napoli, Gruppo
collegato di Salerno 80126 Napoli, Italy}
\email{feoli@unisannio.it}
\author{G.~Lambiase}
\email{lambiase@sa.infn.it}
\author{G.~Scarpetta}
\affiliation{Dipartimento di Fisica "E.R.Caianiello" --
Universit\`a di Salerno,
  84081 Baronissi (SA), Italy
and \\ INFN Sezione di Napoli, Gruppo collegato di Salerno
80126 Napoli, Italy}
\email{scarpetta@sa.infn.it}

\date{\today}
\begin{abstract}
It is shown that the quasi-normal modes arise, in a natural way,
when considering the oscillations in unbounded regions by imposing
the radiation condition at spatial infinity with a complex wave vector $k$.
Hence quasi-normal
modes are not  peculiarities of gravitation problems only (black
holes and relativistic stars).  It is proposed to consider the
space form of the quasi-normal modes with allowance for their time
dependence. As a result, the problem of their unbounded increase
when $r\to \infty$ is not encountered more. The properties of
quasi-normal modes of a compact dielectric sphere are discussed in
detail. It is argued that the spatial form of these modes
(especially so-called surface modes) should be taken into account,
for example, when estimating the potential health hazards due to
the use of portable telephones.
\end{abstract}

 \maketitle
 \section{Introduction}
Quasi-normal modes (qnm) are widely used now in black hole physics
and in relativistic theory of stellar structure (see, for example,
Refs.\ \cite{Nollert,FN,KS}). The corresponding eigenfrequencies
are complex numbers, however it is not due to the dissipative
processes but it is a consequence of unbounded region occupied by
the oscillating system. The latter naturally leads to the energy
loss  due to the wave emission (for example, gravity waves).

In this paper we would like to show that the quasi-normal modes are
not the peculiarities of the gravitational problems only. Actually
they appear, in a natural way, when considering the oscillating
systems unbounded in space. The necessary condition for emergence of
such modes is imposing the radiation condition at spatial infinity on
the field functions. It is this condition that leads to the
characteristic  behaviour of the quasi-normal modes, namely, these solutions
to the relevant equations exponentially decay in time when $t\to
\infty$ and simultaneously they exponentially rise  at spatial infinity $r\to \infty$.
An interesting and physically
motivated example  is
provided here by the oscillations of electromagnetic field
connected with a compact dielectric sphere placed in unrestricted
homogeneous media with a different refraction index or in vacuum.
Taking here  the formal limit $\varepsilon \to \infty$ one passes
to a perfectly conducting sphere ($\varepsilon $ is the refraction
index of the sphere material). In this case the quasi-normal modes
describing the electromagnetic oscillations outside the sphere are
tractable  analytically. We propose to consider the spatial form
of a quasi-normal modes with allowance for their time dependence.
Doing in this way one can escape the  exponential rise of
quasi-normal modes at spatial infinity.

The eigenfrequencies of a dielectric sphere are complex $\omega
=\omega'-i\, \omega''$, where $\omega '$ is the free oscillation
(radian) frequency and $\omega ''$ is its relaxation time. These
modes  can be classified as the
interior  and exterior ones and, at the same time, as volume modes
and surface modes.

In physical applications the surface modes turn out to be
important, for example, when estimating the health hazards due to
the use of portable telephones. The point is the eigenfrequencies
of a dielectric sphere with physical characteristics close to those
of a human head  lay in the GSM 400 MHz frequency band which has
been used in a first generation of mobile phone systems and now is
considered for using again. In this situation one can assume that
the surface modes excited by a  cellular phone will lead to  higher
heat generation in the tissues close to a head surface as compared
with the predictions of routine calculations in this field.

The layout of the paper is as follows. In Sec.\ II we show that
the quasi-normal modes are the eigenfunctions of unbounded
oscillating regions.  As a simple example the quasi-normal modes
of a perfectly conducting sphere are considered. The Sec.\  III is
devoted to the consideration of the quasi-normal modes of a
dielectric sphere. The main features of these modes are revealed and
their classification is presented. The implication of these
quasi-normal modes  for estimation of the health hazards of
portable telephones is considered in Sec.\ IV. In Conclusion
(Sec.\ V) the main results are formulated and their relation to
the general theory of open systems is discussed.

\section{Quasi-normal modes as the eigenfunctions of unbounded oscillating regions}

Here we show in the general case in what way complex frequencies
and quasi-normal modes appear when considering harmonic
oscillations in unbounded regions. Let a closed smooth surface $S$
divides the $d$-dimensional Euclidean space $\mathbb{R}^d$ into a
compact internal region $D_{\text{ in}}$ and noncompact external
region $D_{\text{ex}}$. We consider here a simple scalar wave
equation
\begin{equation}
\label{2-1} \left ( \Delta -\frac{1}{c^2} \frac{\partial
^2}{\partial t^2} \right )u(t, {\textbf x})=0\,{,}
\end{equation}
where $c$ is the velocity of oscillation propagation and $\Delta $
is the Laplace operator in $\mathbb{R}^d$. For harmonic
oscillations
\begin{equation} \label{2-2} u(t,{\textbf x})=e^{-i\omega
t}u({\textbf x}) \end{equation} the wave equation (\ref{2-1}) is
reduced to the Helmholtz equation
\begin{equation} \label{2-3}
(\Delta +k^2)\,
 u(\textbf{x}) =0, \quad
k={\omega}/{c}\,{.} \end{equation}

  The oscillations in the internal region $D_{\text{in}}$ are described by
an infinite countable set of normal modes
\begin{equation}
 \label{2-4}
u_n(t,\textbf{x})=e^{-i\omega _n t}u_n(\textbf{x}), \quad
n=1,2,\ldots .
 \end{equation}
The spatial form of the normal modes (the functions
$u_n(\textbf{x})$) is determined by the boundary conditions which
are imposed upon the function $u(\textbf{x})$ on the internal side
of the surface $S$. These conditions should fit the physical
content of the problem under study. The set of normal modes is a
complete one. Hence any solution of  (\ref{2-3}) obeying relevant
boundary conditions can be expanded in terms of the normal
modes~$u_n(\textbf{x})$.

When considering the oscillations in the external domain $D_{\rm
ex }$ one imposes, in addition to the conditions on the compact
surface $S$, a special requirement concerning the behavior of the
function $u(\textbf{x})$ at large $r\equiv |\textbf{x}|$. In the
classical mathematical physics~\cite{RR} the radiation conditions,
proposed by Sommerfeld~\cite{Sommerfeld,FM}, are used here
\begin{equation}
\label{2-5} \lim_{r\to \infty} r^{\frac{d-1}{2}}u(r)=
\textrm{const}\,{,}\qquad  \lim_{r\to\infty}r^{\frac{d-1}{2}}\left
( \frac{\partial u}{\partial r} -i ku\right )= 0\,{.}
\end{equation}
For real values of the wave vector $k$ (for real frequencies
$\omega$) the solution to Eq.\ (\ref{2-3}), which obeys the radiation
conditions (\ref{2-5}) and reasonable boundary condition on a
compact surface $S$, identically vanishes. In this case the Laplace operator
entering the Helmholtz equation (\ref{2-3}) has no eigenfunctions with
real eigenvalues.

    The physical content of the radiation conditions is very clear.
They select only the oscillations with real frequencies driven by
external sources which are situated in a compact spatial area. From
the mathematical standpoint, these conditions ensure the {\it
uniqueness} of the solutions to {\it inhomogeneous} wave or
Helmholtz equations with external sources on the right-hand side,
when these solutions are considered in the external region $D_{\text{ex}}$ or in
the whole space $D_{\text{in}}+D_{\text{ex}}$  in the case of
compound media.

Here the question arises, how to change minimally the conditions in
the {\it homogeneous} problem at hand in order to get nonzero
solutions, i.e., eigenfunctions in unbounded regions. The energy
conservation law prompts a simple way to construct nonzero solutions
to the homogeneous wave equation (\ref{2-1})  or to the Helmholtz equation
(\ref{2-3}) describing the outgoing waves at spatial infinity, namely,
one has to introduce {\it complex} frequencies $\omega =
\omega'- i\,\omega '',\quad \omega ''>0$. We may hope that in this case
the factor $e^{-\omega ''t}$ will describe the decay of the initial
solutions in time accounting the fact that outgoing waves take away
the energy. In other words, we are dealing here with the radiation
of the energy with the amplitude decaying in time.

  Indeed, if we remove the requirement of reality of the wave vector $k$,
then  the homogeneous wave equation (2.1) and the Helmholtz equation
(2.3) will have nonzero solutions with {\it complex frequencies},
these solutions obeying  the radiation conditions (\ref{2-5}) and a common
boundary condition on a compact surface $S$ (for instance, Dirichlet
or Neumann conditions). In quantum mechanics the radiation condition
with a complex wave number $k$ is known as the Gamov condition which
singles out the resonance states in the spectrum of the Hamiltonian~\cite{Gamov,G1,G2}.

    When introducing the radiation conditions and proving the
respective uniqueness theorem in the text books \cite{RR} only the real
wave vector $k$ is considered. The possibility of existence of
quasi-normal modes with complex frequencies satisfying the radiation
conditions at spatial infinity with a complex $k$ is not mentioned
usually. We are aware only of alone textbook where the
eigenfunctions with complex frequencies are noted in this context.
It is the article written by Sommerfeld in the book~\cite{FM}, where it is
emphasized that the uniqueness in this problem is only up to the
eigenfunctions with complex frequencies, i.e., up to the quasi-normal
modes.

    Thus imposing the radiation conditions  with real $k$ we remove the
quasi-normal modes from our consideration only. However this cannot
prevent the excitation of these modes in the real physical problem.
Hence, when dealing with systems unbounded in space (open systems)
one has always to investigate the consequences of qnm excitation. An
example of such a problem will be considered in Sec.~III.

As a very simple and physically motivated example of quasi-normal modes we
consider here the oscillations of electromagnetic field outside a
perfectly conducting sphere of radius $a$. In this case the
electric and magnetic fields are expressed in terms of two scalar
functions $f_{kl}^{\rm{TE}}(r)$ and $f_{kl}^{\rm{TM}}(r)$ (Debye
potentials \cite{Stratton}) which are the radial parts of the
solutions to the scalar wave equation (\ref{2-1}). Outside the
perfectly conducting sphere placed in vacuum the solution to the
Helmholtz equation (\ref{2-3}) obeying the radiation conditions
(\ref{2-5}) has the form $(d=3)$
\begin{equation}
\label{2-6} f_{kl}(r)=C\, h^{(1)}_l\left ( \frac{\omega}{c}\,r
\right ){,}\quad r>a\,{,}
\end{equation}
where $ h^{(1)}_l(z)$ is the spherical Hankel function of the
first kind \cite{AS}. At the surface of perfectly conducting
sphere the tangential component of the electric field should
vanish. This leads to the following frequency equation for
TE-modes
\begin{equation}
\label{2-7} h^{(1)}_l\left ( \frac{\omega}{c}\,a \right
)=0{,}\quad l\geq 1
\end{equation}
and for TM-modes
\begin{equation}
\label{2-8} \frac{d}{d r}\left ( r \,h^{(1)}_l\left (
\frac{\omega}{c}\,r \right ) \right )=0, \quad r=a,\quad l\geq
1\,{.}
\end{equation}
The spherical Hankel function $h^{(1)}_l(z)$ is $e ^{i z}$
multiplied by the polynomial in $1/z$  of a finite order
\cite{AS}. Hence frequency equations (\ref{2-7}) and (\ref{2-8})
have a finite number of roots which are in the general case
complex numbers. For $l=1$ (the lowest oscillations) Eqs.\
(\ref{2-7}) and (\ref{2-8}) assume  the form $(z=a\,\omega/c)$
\begin{eqnarray}
h^{(1)}_l(z)&=&-\frac{1}{z}\,e^{i z}\left ( 1+\frac{i}{z} \right
)=0\quad (\text{TE modes}),
 \label{2-8a} \\
\frac{d }{d z}\left (z\, h^{(1)}_l(z)\right
)&=&-\frac{i}{z^2}\,e^{i z}\left ( z^2+i z-1 \right )=0\quad
(\text{TM modes})\,{.}
 \label{2-9}
\end{eqnarray}
Thus the lowest eigenfrequencies are
\begin{eqnarray}
\frac{\omega}{c}&=&-\frac{i}{a}\quad (\mbox{TE modes})\,{,}
\label{2-10}
\\
\frac{\omega}{c}&=&-\frac{1}{2a}(i\pm\sqrt 3)\quad (\text{TM
modes})\,{.} \label{2-11}
\end{eqnarray}

The complex eigenfrequencies lead to a specific time and spatial
dependence of the  respective natural modes and ultimately of the
electromagnetic fields. So, with allowance of (\ref{2-10}), we
obtain
\begin{equation}
\label{2-12} e^{-i \omega t}f_{k1}^{\text{TE}}(r)=-i
\,C\,\frac{a}{r}\,e^{(r-ct)/a}\left ( 1-\frac{a}{r} \right ),
\quad r\geq a\,{.}
\end{equation}
Thus, the eigenfunctions are exponentially going down in time and
exponentially going up when $r$ increases. Such a time and spatial
behaviour is a direct consequence of radiation conditions (\ref{2-5}) and it is
typical for eigenfunctions describing oscillations in
external unbounded regions, the physical content and details of
oscillation process  being irrelevant. The eigenfunctions
corresponding to complex eigenvalues are called {\it quasi-normal
modes} keeping in mind their unusual properties~\cite{Nollert}.
The physical origin of such features is obvious, in fact  we are
dealing here with {\it open systems} in which the energy  can be
radiated to infinity. Therefore in open systems field cannot
acquire a stationary state.

 Quasi-normal modes do not obey the standard completeness condition
and the notion of norm cannot be defined for them~\cite{Nollert}.
Therefore these eigenfunctions cannot be used for expansion of the
classical field with the aim to quantize it and to introduce the
relevant Fock operators. The treatment of these problems can be found,
for example, in~\cite{Ching,Leung,Chang}.

It is worthy to investigate the spatial form of the of quasi-normal
modes with allowance for their time dependence. Indeed, these
solutions have the character of propagating waves that are
eventually going to spatial infinity. Let us take the point which
is sufficiently far from  the region with nontrivial dynamics in
the system under study. Obviously, it has sense to say about the
value of the quasi-normal mode at a given point only after arrival
at this point of the wave described by this  mode. The maximal
value of the quasi-normal mode  is observed just at the moment of
its arrival at this point. At the later moments the quasi-normal
mode is dumping due to its characteristic time dependence. Indeed,
taking into account all this we obtain for the maximal observed
value of the quasi-normal mode (\ref{2-12}) the following
physically acceptable expression
\begin{equation}
\label{2-13}\left ( e^{-i \omega t}f_{k1}^{\text{TE}}(r)\right
)_{\text{max-obs}}=-i \,C\,\frac{a}{r}\left ( 1-\frac{a}{r} \right
), \quad r\geq a\,{.}
\end{equation}
Thus, in our consideration the problem of unbounded
(exponential) rising of quasi-normal modes, when $r\to \infty$, does
not arise.

In order to associate with resonance phenomenon a single square
integrable eigenfunction, rather sophisticated methods are used, for
example, complex scaling~\cite{scaling} (known also as the
complex-coordinate method or as the complex-rotational method).

 The necessary condition for
appearing the quasi-normal modes is imposing the radiation
conditions at spatial infinity on the field functions. When other
conditions are used at spatial infinity, the quasi-normal modes do
not arise. For example, if we demand that the solution to the wave
equation (\ref{2-1}) becomes the sum of incoming and outgoing
waves when $r\to \infty$ then the spectrum of the variable $k^2$
in the Helmholtz equation (\ref{2-3}) will be positive and
continuous.

\section{Quasi-normal modes  of a dielectric sphere}

 The same situation, with regard to quasi-normal modes, takes
place when we consider the oscillations of compound unbounded
media. In this case in both the regions $D_{\text{in}}$ and
$D_{\text{ex}}$ the wave equations are defined
\begin{eqnarray}
\left ( \Delta -\frac{1}{c_{\text{in}}^2}\frac{\partial }{\partial
t^2} \right ) u_{\text{in}}(t,\mathbf{x}) &=& 0, \quad
\mathbf{x}\in D_{\text{in}}\,{,} \label{2-14}\\ \left ( \Delta
-\frac{1}{c_{\text{ex}}^2}\frac{\partial }{\partial t^2} \right )
u_{\text{ex} }(t,\mathbf{x}) &=& 0, \quad \mathbf{x}\in D_{\text{
ex}} \label{2-15}
\end{eqnarray}
with the matching conditions at the interface $S$, for example, of
the following kind
\begin{eqnarray}  u_{\text{ in}}(t,\mathbf{x})&=&u_{\text{ex}}(t,
\mathbf{x}){,} \label{2-16}
\\
\lambda_{\text{ in}}\frac{\partial u_{\text{in
}}(t,\mathbf{x})}{\partial n_{\text{in
}}(\mathbf{x})}&=&\lambda_{\text{ ex}}\frac{\partial u_{\rm
ex}(t,\mathbf{x})}{\partial n_{\text{ ex }}(\mathbf{x})}, \quad
\mathbf{x}\in S\,{,} \label{2-17}
\end{eqnarray}
where $n_{\text{in}}(\textbf{x})$ and $n_{\text{ ex}}(\textbf{x})$
are the normals to the surface $S$ at the point $\textbf{x}$ for
the regions $D_{\text{in}}$ and $D_{\text{ex}}$, respectively. The
parameters $c_{\text{in}}$, $c_{\text{ex}}$, $\lambda
_{\text{in}}$, and $\lambda _{\text{ex}}$ specify the material
characteristics of the media. At the spatial infinity the solution
$u_{\text{ex}}(t, \mathbf{x})$ should satisfy the radiation
conditions (\ref{2-5}). For real $k$ we again have only zero
solution in this problem, both functions $u_{\text{in}}(t,
\mathbf{x})$ and $u_{\text{ex}}(t, \mathbf{x})$ vanishing.
However, the wave equations (\ref{2-14}) and (\ref{2-15}) have
nonzero solutions with complex frequencies, i.e.\ quasi-normal
modes, which satisfy the matching conditions (\ref{2-16}) and
(\ref{2-17}) at the interface $S$ and radiation conditions
(\ref{2-5}) at  spatial infinity. It is important, that the
frequencies  of oscillations in internal ($D_{\text{in}}$) and
external ($D_{\text{ex}}$) regions are the same. A typical example
here is the complex eigenfrequencies of a dielectric sphere. This
problem has been investigated by Debye in his PhD thesis concerned
with the light pressure on a material particles~\cite{Debye}.

Let us consider a sphere of radius $a$, consisting of a
material which is characterized by permittivity $\varepsilon_1 $
and permeability $\mu_1$. The sphere is assumed to be placed in an
infinite medium with permittivity $\varepsilon_2 $ and
permeability $\mu_2$. In the case of spherical symmetry the
solutions to Maxwell equations are expressed in terms of two
scalar Debye potentials $\psi$ (see, for example, textbooks
\cite{Stratton,Jackson}):
\begin{eqnarray}
\mathbf{E}^{\text{TM}}_{lm}&=&\bm{\nabla} \times
\bm{\nabla}\times(\mathbf{r}\psi^{\text{TM}}_{lm}),\quad
\mathbf{H}^{\text{TM}}_{lm}=-i\,\omega \,\bm{\nabla} \times
(\mathbf{r}\psi^{\text{TM}}_{lm})\quad (\text{E-modes}), \nonumber
\\ \mathbf{E}^{\text{TE}}_{lm}&=&i\,\omega \,\bm{\nabla} \times
(\mathbf{r}\psi^{\text{TE}}_{lm}),\quad
\mathbf{H}^{\text{TE}}_{lm}=\bm{\nabla} \times
\bm{\nabla}\times(\mathbf{r}\psi^{\text{TE}}_{lm})\quad
(\text{H-modes})\,{.} \label{3-1}
\end{eqnarray}
These potentials obey the Helmholtz equation and have the
indicated angular dependence
\begin{equation}
\label{3-2} \left ( \mathbf{\nabla}^2+k^2_i\right
)\psi_{lm}=0,\quad
k_i^2=\varepsilon_i\,\mu_i\,\frac{\omega^2}{c^2}, \quad i=1,2\quad
(r\neq a); \quad \psi_{lm}(\mathbf{r})=f_l(r)Y_{lm}(\Omega)\,{.}
\end{equation}

Equations  (\ref{3-2}) should be supplemented by the boundary conditions at
the origin, at the sphere surface, and at infinity. In order for the
fields to be finite at $r=0$ the Debye potentials should be regular
here. At the spatial infinity we impose the radiation conditions with
the goal to find the spectrum of eigenfunctions with complex frequencies
(quasi-normal modes in the problem at hand). At the sphere surface
the standard matching conditions for electric and magnetic fields
should be satisfied~\cite{Stratton}.

In view of all this the Helmholtz equation (\ref{3-2}) becomes now
the spectral problem for the Laplace operator
multiplied by the discontinuous factor $-1/(\varepsilon
(r)\,\mu(r))$
\begin{equation}
\label{3-2a}-\, \frac{1}{\varepsilon (r)\,\mu(r)}\Delta
\,\psi_{\omega l m}(r)=\frac{\omega^2}{c^2}\,\psi_{\omega l m}(r),
\quad r\neq a\,{,}
\end{equation}
where
\[
\varepsilon (r)\,\mu(r)=
\begin{cases}
\varepsilon_1\,\mu_1\,{,}&r<a \,{,}\\
\varepsilon_2\,\mu_2\,{,}&r>a \,{.}
\end{cases}
\]
In this problem the spectral parameter is $\omega^2/c^2$.

In order to obey the boundary conditions at the origin and at
spatial infinity formulated above, the solution to the spectral
problem (\ref{3-2a}) should have the form
\begin{equation}
\label{3-2b} f_{\omega l}(r)=C_1\,j_l(k_1r)\,{,} \quad r<a,\quad
f_{\omega l}(r)=C_2\,h^{(1)}_l(k_2r)\,{,} \quad r>a\,{,}
\end{equation}
where $j_l(z)$ is the spherical Bessel function and $h_l^{(1)}(z)$
is the spherical Hankel function of the first kind \cite{AS}, the
latter obeys the radiation conditions (\ref{2-5}).

Now we address the matching conditions at the sphere surface.
By making use of Eqs.\ (\ref{3-1}) we can write, in an explicit
form, the radial (r) and tangential (t) components of electric and
magnetic fields in the case of spherical symmetry. For TE-modes
these equations read
\begin{eqnarray}
E^{\text{TE}}_{klm,\text{r}}&=&0\,{,} \label{a-20}\\
E^{\text{TE}}_{klm,\text{t}}&=&a_{lm}(k)\,f_{kl}^{\text{TE}}(r)\,\mathbf{X}_{lm}\,{,}
\label{a-21}\\ H^{\text{TE}}_{klm,\text{r}}&=&\frac{1}{kr}\left (
\frac{\varepsilon}{\mu} \right
)^{1/2}\sqrt{l(l+1)}\,a_{lm}(k)f_{kl}^{\text{TE}}(r)\,Y_{lm} \,{,}
\label{a-22}\\ H^{\text{TE}}_{klm,\text{t}}&=&\frac{i}{kr}\left (
\frac{\varepsilon}{\mu} \right
)^{1/2}a_{lm}(k)\,\frac{d}{dr}\left( rf_{kl}^{\text{TE}}(r)
\right)\,\mathbf{X}_{lm}^\perp\,{,} \label{a-23}
\end{eqnarray} and the same for the TM-modes
\begin{eqnarray}
E^{\text{TM}}_{klm,\text{r}}&=&-\frac{1}{kr}\left (
\frac{\mu}{\varepsilon} \right
)^{1/2}\sqrt{l(l+1)}\,b_{lm}(k)f_{kl}^{\text{TM}}(r)\,Y_{lm} \,{,}
\\ E^{\text{TM}}_{klm,\text{t}}&=&-\frac{i}{kr}\left (
\frac{\mu}{\varepsilon} \right
)^{1/2}b_{lm}(k)\,\frac{d}{dr}\left( rf_{kl}^{\text{TM}}(r)
\right)\,\mathbf{X}_{lm}^\perp\,{,} \label{a-25}\\
H^{\text{TM}}_{klm,\text{r}}&=&0\,{,} \label{a-26}\\
H^{\text{TM}}_{klm,\text{t}}&=&b_{lm}(k)\,f_{kl}^{\text{TM}}(r)\,\mathbf{X}_{lm}\,{.}
\label{a-27}
\end{eqnarray}
Here $\mathbf{X}_{lm}$ are the vector spherical
harmonics~\cite{Jackson}
\begin{equation}
\label{vsh}
\mathbf{X}_{lm}(\theta,\phi)=\frac{\mathbf{L}\,Y_{lm}(\theta,\phi)}{\sqrt{l(l+1)}}\,{,}\quad
l\geq1\,{,}
\end{equation}
where $\mathbf{L}$ is the angular momentum operator
\[
\mathbf{L}=-i\,(\mathbf{r}\times \bm{\nabla})\,{.}
\]
The vector spherical harmonic  $\mathbf{X}_{lm}^{\perp}$ is
obtained from $\mathbf{X}_{lm}$ after rotation  by the angle
$\pi/2$ around the normal $\mathbf{n}=\mathbf{r}/r$. From Eqs.\
(\ref{a-20}) -- (\ref{a-27}) it follows, in particular, that the
tangential components of electric field in TE- and TM-modes are
orthogonal each other and the same holds for the magnetic field.
It implies that the matching conditions on the sphere surface do not
couple TE- and TM-modes.

At the sphere surface the tangential components of electric and
magnetic fields are continuous (see Eqs.\ (\ref{a-20}) -- (\ref{a-27})).
As a result, the eigenfrequencies
of electromagnetic field for this configuration are determined
\cite{Stratton} by the frequency equation for the TE-modes
\begin{equation}
\label{3-3} \Delta^{\text{TE}}_l(a\omega)\equiv
\sqrt{\varepsilon_1\mu_2}\,\hat j_l'(k_1a)\,\hat h_l(k_2a)-
\sqrt{\varepsilon_2\mu_1}\,\hat j_l(k_1a)\,{\hat h_l}'(k_2a)=0\,{}
\end{equation}
and by the analogous equation for the TM-modes
\begin{equation}
\label{3-4} \Delta^{\text{TE}}_l(a\omega)\equiv
\sqrt{\varepsilon_2\mu_1}\,\hat j_l'(k_1a)\,\hat h_l(k_2a)-
\sqrt{\varepsilon_1\mu_2}\,\hat j_l(k_1a)\,{\hat
h_l}'(k_2a)=0\,{,}
\end{equation}
where $k_i=\sqrt{\varepsilon_i\mu_i}\,\omega/c,\quad i=1,2$ are
the wave numbers inside and outside the sphere, respectively, and
$\hat j_l(z)$ and $\hat h_l(z)$ are the Riccati-Bessel
functions~\cite{AS}
\begin{equation}
\label{3-5} \hat j_l(z)=z\,j_l(z)=\sqrt{\frac{\pi
z}{2}}\,J_{l+1/2}(z)\,{,}\quad \hat
h_l(z)=z\,h_l^{(1)}(z)=\sqrt{\frac{\pi
z}{2}}\,H^{(1)}_{l+1/2}(z)\,{.}
\end{equation}
In Eqs.\ (\ref{3-3}) and (\ref{3-4}) the orbital momentum $l$
assumes the values $1,2,\ldots $, and prime stands for the
differentiation with respect of the arguments $k_1a$ and $k_2a$ of
the Riccati-Bessel functions.

The frequency equations for a dielectric sphere of permittivity
$\varepsilon $ placed in vacuum follow from Eqs.\ (\ref{3-3}) and
(\ref{3-4}) after putting there
\begin{equation}
\label{3-12} \varepsilon_1=\varepsilon, \quad
\varepsilon_2=\mu_1=\mu_2=1\,{.}
\end{equation}
The roots of these equations have been studied in the Debye paper
\cite{Debye} by making use of an approximate  method. As the
starting solution the eigenfrequencies of a perfectly conducting
sphere were used. These frequencies are different for
electromagnetic oscillations inside and outside  sphere. Namely,
inside  sphere they are given by the roots of the following
equations $(l\geq 1)$
\begin{eqnarray}
j_l\left ( \frac{\omega}{c}\,a \right)&=&0 \quad
(\text{TE-modes})\,{,} \label{3-13}\\ \frac{d}{dr}
\left(r\,j_l\left ( \frac{\omega}{c}\,a \right)\right)&=&0\,{,}
\quad r=a \quad (\text{TM-modes})\,{,} \label{3-14}
\end{eqnarray}
while outside sphere they are determined by Eqs.\ (\ref{2-7}) and
(\ref{2-8}). The frequency equations for perfectly conducting
sphere (\ref{2-7}), (\ref{2-8}) and  (\ref{3-13}), (\ref{3-14})
can be formally derived  by substituting (\ref{3-12}) into
frequency equations (\ref{3-3}) and (\ref{3-4}) and taking there
the limit $\varepsilon \to \infty$.

 Approximate calculation of the eigenfrequencies of a dielectric
sphere without using computer \cite{Debye} didn't allow one to reveal
the characteristic features of the respective eigenfunctions
(quasi-normal modes). The computer analysis of this spectral problem
was accomplished in the work~\cite{Gastine} where the experimental
verification of the calculated frequencies was accomplished also by
making use of radio engineering measurements.

These studies enable one to separate all the dielectric sphere modes into the
{\it interior}  and {\it exterior} modes and, at the same time, into
the {\it volume} and {\it surface} modes. It is worth noting that all
the eigenfrequencies are complex
\begin{equation}
\label{3-15} \omega=\omega '-i\,\omega''\,{.}
\end{equation}
Thus we are dealing with "leaky modes".

The classification of the modes as the interior and exterior  ones
relies on the investigation of the behaviour of a given
eigenfrequency in the limit $\varepsilon \to \infty$. The modes are
called ''interior'' when the product $k\,a= \sqrt
{\varepsilon}\,\omega \,a/c$ remains finite in the limit
$\varepsilon \to \infty$, provided the imaginary part of the
frequency ($\omega ''$) tends to zero. The modes are referred to
as ''exterior'' when the product $k\,a/\sqrt{\varepsilon}=
\omega\,a/c$ remains finite with growing $\omega''$. In the first
case the frequency equations for a dielectric sphere (\ref{3-3}) and
(\ref{3-4}) tend to Eqs.\ (\ref{3-13}) and (\ref{3-14}) and in the
second case they tend to Eqs.\ (\ref{2-7}) and  (\ref{2-8}). The
order of the root obtained will be denoted by the index $r$ for
interior modes and by $r'$ for exterior modes. Thus
$\text{TE}_{\,lr}$ and $\text{TM}_{\,lr}$  denote the interior TE-
and TM-modes, respectively,  while $\text{TE}_{\,lr'}$ and
$\text{TM}_{\,lr'}$ stand for the exterior TE- and TM-modes.

For fixed $l$ the number of the modes of exterior type is limited
because the frequency equations for exterior oscillations of a
perfectly conducting sphere (\ref{2-7}) and  (\ref{2-8}) have
finite number of solutions (see the preceding Section). In view of
this, the number of exterior TE- and TM-modes is given by the
following rule. For even $l$ there are $l/2$ exterior TE-modes and
$l/2$  exterior TM-modes, for odd $l$ the number of the modes
$\text{TE}_{\,l\,r'}$ is $(l+1)/2$ and the number of the modes
$\text{TM}_{\,l\,r'}$ equals $(l-1)/2$.

An important parameter is the $Q$ factor
\begin{equation}
\label{3-16} Q_{\text{rad}}=\frac{\omega'}{2\omega''}=
2\,\pi\,\frac{\text{stored energy}}{\text{radiated energy per
cycle}}\,{.}
\end{equation}
For exterior modes the value of $Q_{\text{rad}}$ is always less than 1, hence
these modes can never be observed as sharp resonances. At the same time for
$\varepsilon $ greater than 5, the $Q_{\text{rad}}$ for interior modes is
greater than 10 and it can reach very high values when
$\varepsilon \to \infty$.

For physical implications  more important is the classification in
terms of {\it volume} or {\it surface} modes according to whether
$r> l$ or $l> r$.  For volume modes the electromagnetic energy
is distributed in the whole volume of the sphere while in the case
of surface modes the energy is concentrated in the proximity of
the sphere surface. The exterior modes are the first roots of the
characteristic equations and it can be shown that they are always
surface modes.
\noindent
\begin{figure}[th]
\noindent \centerline{
\includegraphics[width=75mm]{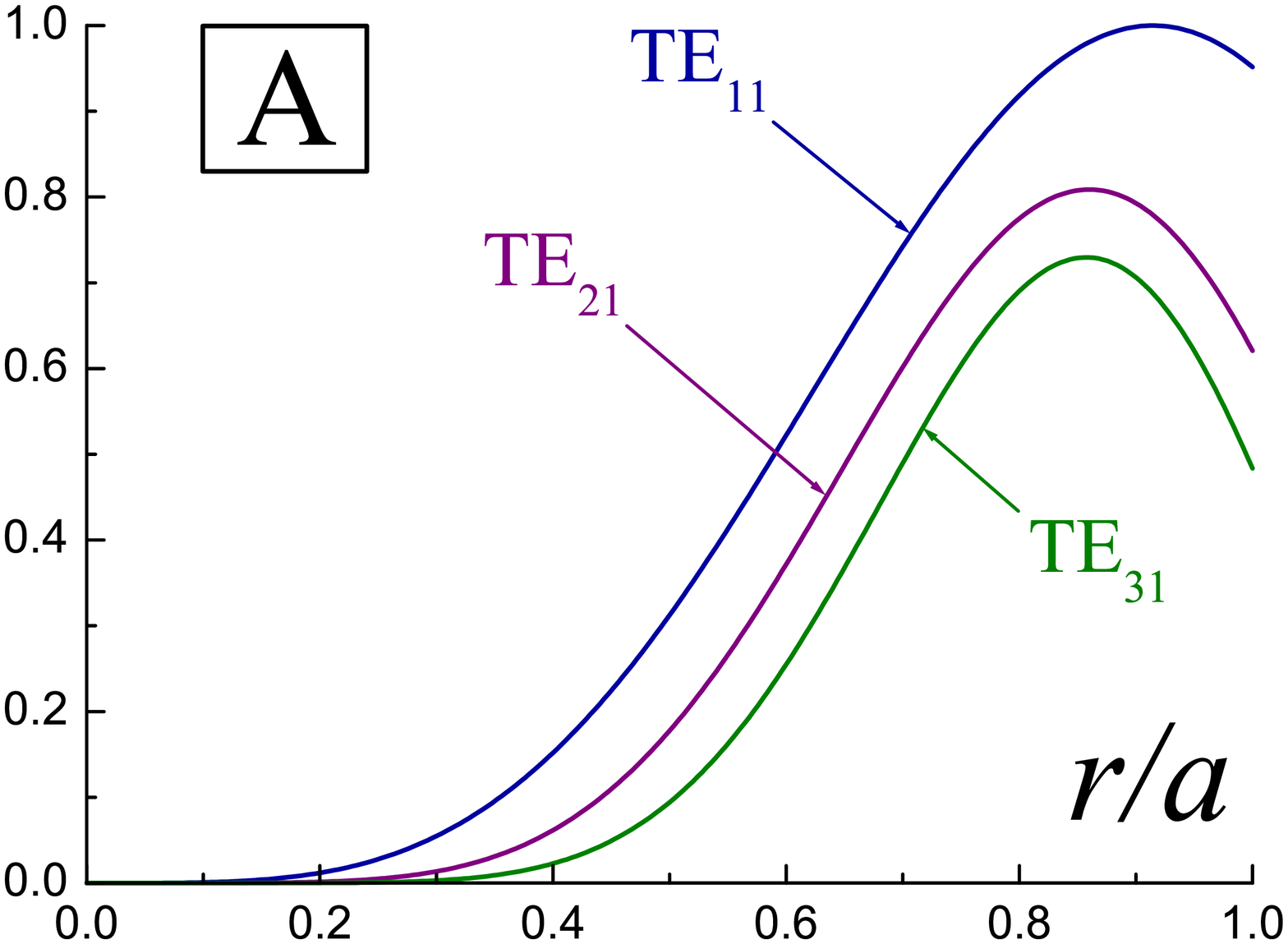}
\hspace{10mm}
\includegraphics[width=75mm]{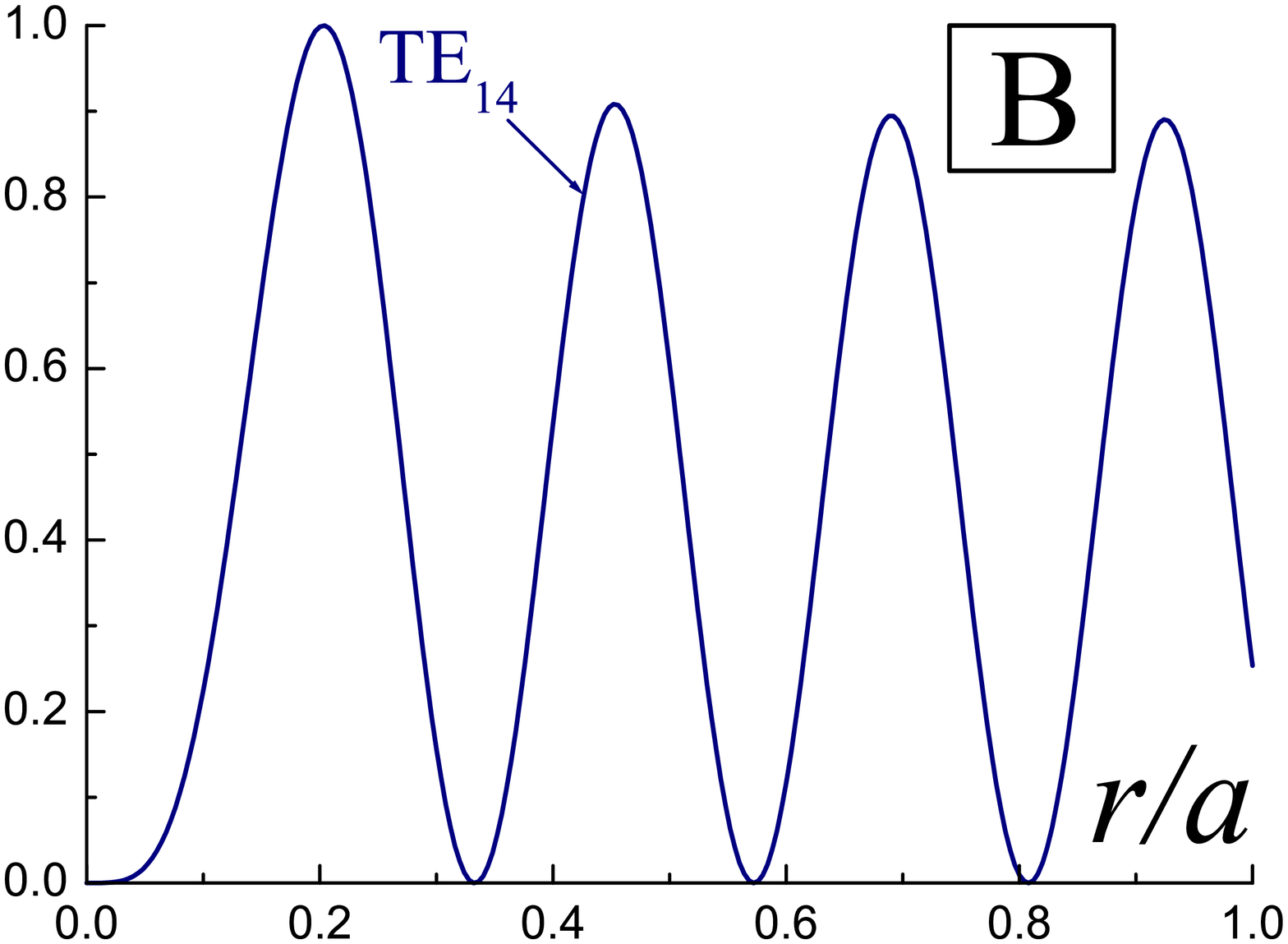}}
\caption{Electric energy density $r^2\,E_{\text{t}}^2$ for the
surface (A) and volume (B) TE-modes of a dielectric sphere with
$\varepsilon =40$ placed in vacuum.} \label{Plot:s-v-modes}
\end{figure}

Figure 1 shows a typical spatial behaviour of the surface and
volume modes of a dielectric sphere.

Thus a substantial part of the sphere modes (about one half) belong to the interior
surface modes. It is
important that respective frequencies are the {\it first} roots of
the characteristic equations.

In order to escape the confusion, it is worth noting here that the
surface modes in the problem in question obey the same boundary
conditions at the sphere surface  and when $r\to \infty$ as the
volume modes do. Hence,  these surface modes cannot be classified as the
evanescent surface waves propagating along the interface  between
two media (propagating waves along dielectric waveguides~\cite{Jackson}, surface
plasmon waves on the interface between metal bulk and adjacent
vacuum~\cite{Raether,BPN} and so on).  When describing the evanescent waves one imposes
the requirement of their exponential decaying away from interface
between two media. In this respect the evanescent surface wave
differ from the modes in the bulk.

\section{Implication of QNM of a dielectric sphere
for estimation of the health hazard of portable telephones}

Here we shall argue that the features of the quasi-normal modes of a
dielectric sphere (namely, existence of surface and volume modes)
should be taken into account, in
particular, when estimating the potential  health hazards due to
the use of the cellular phones. The safety guidelines in this
field~\cite{Health} are based on the findings from animal
experiments that the biological hazards due to radio waves result
mainly from the temperature rise in tissues\footnote{In principle,
non-ionizing radiation can lead also to other effects in
biological tissues~\cite{Sernelius}.} and a whole-body-averaged
specific absorption rate (SAR) below 0.4~W/kg is not hazardous to
human health. This corresponds to a limits on the output from the
cellular phones (0.6~W at 900~MHz frequency band and 0.27~W at
1.5~GHz frequency band). Obviously, the {\it local} absorption
rate should be also considered especially in a human
head~\cite{WF}.

In such studies the following point should be
taken into account. The parts of human body (for example, head)
posse the  eigenfrequencies of electromagnetic oscillations like
any compact body. In particular, one can anticipate that the
eigenfrequencies of human head are close to those of a dielectric
sphere with radius $a\approx 8$~cm and permittivity $\varepsilon
\approx 40$ (for human brain $\varepsilon =44.1$ for  900~MHz and
$\varepsilon =42.8$ for  1.5~GHz \cite{WF}).
Certainly, our model is very rough, however for the evaluation of the order of
the effect anticipated (see below)
it is sufficient.
By making use of the
results of calculations conducted in the work \cite{Gastine} one
can easily obtain the eigenfrequencies of a dialectic sphere with
the parameters  mentioned above. For $\text{TE}_{l1}$ modes with
$l=1,2,3$ we have, respectively, the following frequencies:
280~MHz, 420~MHz, and 545~MHz. For $\text{TM}_{l1}$ modes
with $l=1,2,3$ the resonance frequencies are 425~MHz, 540~MHz, and
665~MHz. The imaginary parts of these eigenfrequencies are very
small so the $Q$ factor in Eq.\ (\ref{3-16}) responsible for radiation
is greater than 100.

These eigenfrequencies belong to a new GSM 400 MHz frequency band
which is now being standardized by the European Telecommunications
Standards Institute.  This band was primarily used in Nordic
countries, Eastern Europe, and Russia in a first generation of
mobile phone system prior to the introduction of GSM.

    Due to the Ohmic losses the resonances of a dielectric sphere
 in question are in fact broad, overlapping and,
as the result, they cannot be manifested separately.
Indeed, the electric conductance $\sigma$ of the human brain is rather substantial.
According to the data  presented in Ref.~\cite{WF} $\sigma\simeq 1.0$~S/m.
The eigenfrequencies  of a dielectric dissipative sphere with allowance for
a finite conductance $\sigma $ can be found in the following way. As known \cite{LL}
the effects of $\sigma $ on electromagnetic processes in a media possessing
a common real dielectric constant $\varepsilon$ are described by a complex
dielectric constant $\varepsilon_{\text{diss}}$ depending on frequency
\begin{equation}
\label{4-1}
\varepsilon_{\text{diss}} =\varepsilon +i\frac{4\pi \sigma}{\omega}\,{.}
\end{equation}
The eigenfrequencies $\omega$,  calculated for a real $\varepsilon $,
are related to eigenfrequencies $\omega_{\text{diss}}$ for $\varepsilon_{\text{diss}}$
by the formula~\cite{LL}
\begin{equation}
\label{4-2}
\omega_{\text{diss}} =\frac{\omega}{\sqrt{\varepsilon_{\text{diss}} }}\simeq \omega
-2\pi\, i \,\frac{\sigma}{\varepsilon}\,{.}
\end{equation}
The corresponding factor $Q_{\text{diss}}$ is
\begin{equation} \label{4-3} Q_{\text{diss}}=\frac{\omega'_{\text{diss}}}{2
\omega''_{\text{diss}}}\simeq
\frac{\varepsilon \, \omega}{4\pi\, \sigma}\,{.}
\end{equation}
Substituting in this equation the values
$\omega /2\pi =0.5\cdot 10^9\;\text{Hz}, \quad \varepsilon =40,
\quad \sigma = 1\,\text{S/m}=9\cdot 10^9\; \text{s}^{-1}$ one finds
\begin{equation}
\label{4-4}
Q_{\text{diss}}\simeq\frac{20}{18}\simeq 1\,{.}
\end{equation}

Thus the real
spectrum of electromagnetic oscillations in the problem under study
is practically a continuous band around the frequency 400 MHz. The
radiation of the cellular telephone with frequency laying in this
band, will excite (practically with the same amplitudes) all the
neighbouring modes of a dissipative dielectric sphere. In order to
get the upper bound for the anticipated effect (see below) we assume
that the number of excited modes is sufficiently large so that
the half of these modes are surface modes.\footnote{As it was
shown in preceding Section this relation between the number of the
surface and volume modes holds only for the spectra as a whole.}
Thus one can expect that the resulting spatial configuration of
electric and magnetic fields inside a dielectric sphere with Ohm
losses will follow, to some extent, the spatial behaviour of the
relevant natural modes of the sphere, volume and surface ones. It is
obvious that due to excitation of surface modes the maximum values
of electric and magnetic fields inside dissipative sphere will be
shifted to its outer part $r> a/2$.

When assuming the total number of the surface modes to be the same
as those for the  volume modes and consequently it is equal to a half
of all the dielectric sphere modes, then one can anticipate that the
temperature rise in the head tissues close to head surface may be by
a factor 1.5 higher  in comparison with the standard calculations
using the  numerical methods without  special allowance for the spatial
behaviour  of the relevant natural modes.

However the numerical methods used for estimation of the
temperature rise in human tissues due to the radio frequency
irradiation  do not take into account this effect. Indeed, such
calculations (see, for example,  paper~\cite{WF}) are carried out
in two steps. First the electric and magnetic fields inside the
human body are calculated by solving the Maxwell equations with a
given source (antenna of a portable telephone). The electric field
gives rise to conduction currents with the energy dissipation
rate $\sigma \,E^2/2$, where $\sigma $ is a conduction constant.
In turn it leads to the temperature rising. The second step is the
solution of the respective heat conduction equation (or more
precisely, bioheat equation~\cite{WF}) with found local heat
sources $\sigma \,E^2/2$ and with allowance for all the possible
heat currents. Hence, for this method the  distribution of
electric field inside the head is of primary importance.
The spatial behaviour of the
eigenfunctions characterize the system as a whole, and  these
properties cannot be taken into account by local methods for
calculating the solution to partial differential equations (in
our case, to the Maxwell equations).

\section{Conclusion}
We have shown that such different, at first glance, notions as
quasi-normal modes in black hole physics, Gamov states in quantum
mechanics, and quasi-bound states in the theory of open
electromagnetic resonators~\cite{Vaynstayn} have the same origin, namely, all these
are the eigenmodes of oscillating unbounded domains. By making use of
a simple but physically motivated example of electromagnetic
oscillations outside a perfectly conducting sphere, which admits
analytical treatment, we have easily shown the main features  of such
oscillations, namely, their exponential decaying in time and,
simultaneously, their exponential grows at spatial infinity. These
properties are a direct consequence of the radiation condition which
is met by qnm at spatial infinity. It is shown also that the
exponential rising of qnm at infinity is not observable because the
time dependence of qnm should be taken into account here. In the
considered example the qnm are the outgoing spherical waves for
$r>a$. This point is disregarded in all the attempts to treat qnm
mathematically, in particular, to formulate the completeness
condition and the relevant expansions in terms of qnm. We have
considered the qnm modes in the problem without inhomogeneous
potential (instead of the potential the boundary conditions at the
surface of the sphere are introduced). It enables us to infer
the conclusion stated above clear and easy.

 It is argued also that imposing the standard radiation conditions with
a real wave vector $k=\omega /c$ does not prevent us from the
necessity to investigate the eigenmodes with complex $\omega'$s,
i.e., qnm in a given problem.

 The importance of this is demonstrated by
investigating the role of the qnm in the problem of estimating the
potential health hazards due to the use of portable telephones. The
general analysis of the qnm spectra of a dielectric sphere with
allowance for the dissipative processes enables us to estimate
quantitatively the expected effect, namely, a possible temperature
rise in the tissues laying in the outer part of the human head may
be 1.5 times greater as compared with the inner part of the brain.

    The predicted effect is, in some sense, analogous to the usual (but
weakly manifested here) skin-effect. It is not surprising because
the substantial conductivity of the sphere material plays a
principal part in our consideration. Due to this conductivity the
individual resonances of a sphere become very overlapping and, being
not observed separately, many of them (volume and surface ones) are
excited by the cellular telephone irradiation with the frequency in
this band. In view of different spatial behaviour, the surface and
volume modes will lead to different spatial distributions of the net
heat inside the sphere (and also inside a human head).

When  we have the quasi-normal modes instead the usual normal
modes it implies that we are dealing with the open
systems~\cite{Open}. Open systems admit a dual description: on
the one hand, they can be considered from the ``inside'' point of
view, treating the coupling to the environment as a -- not
necessarily small -- perturbation. From this point of view, one
can study the (discreet) eigenvalues of the system, the width of
resonances and the resulting decay properties~\cite{Ching,Leung}. On the other hand,
open systems allow to take the ``outside'' point of view,
considering the system as a perturbation of the environment. The
typical quantity to be investigated from this point of view is the
{\it scattering matrix} ($S$-matrix), i.e.\ the amplitude for
passing from a given incoming field configuration  to a certain outgoing
configuration as a function of energy~\cite{Vaynstayn}.

The registration of the quasi-normal modes of a black hole is
considered now as a possible way to detect this object
\cite{Nollert,KS}. In this connection it is surpassing that till
now the quasi-normal modes have not been used in an analogous
acoustic problem, i.e., for description of sound generation by ringing
body~\cite{Raylaygh,Lamb,Morse}. At the same time one
can find in literature the statement (without proving)
that we hear ringing quasi-normal modes of a bell when we hear the
bell sound~\cite{Moss}.

\begin{acknowledgments}
 This paper was completed during the visit of on of the authors
(VVN) to Salerno University. It is his pleasant duty to thank G.\
Scarpetta and G.\ Lambiase for the kind hospitality extended to
him. VVN was supported in part by the Russian Foundation for Basic
Research (Grant No.\ 03-01-00025). The financial support of INFN
is acknowledged. The authors are indebted to A.V.~Nesterenko for
preparing the figure.
\end{acknowledgments}

\end{document}